\magnification = \magstep1
\pageno=0
\hsize=15.0truecm
\hoffset=1.0truecm
%
\vsize=23.5truecm

\output={\plainoutput}
\pretolerance=3000
\tolerance=5000
\hyphenpenalty=10000    
\newdimen\digitwidth
\setbox0=\hbox{\rm0}
\digitwidth=\wd0
\def\cl{\centerline}

\def\vs{\vskip 11pt}

\font\sc=cmr7

\def\ia{(i/\kern-2pt {\sc A})}
\def\ja{(j/\kern-2pt {\sc A})}
\ \
\vs\vs\vs
\vs
\vs
\vs
\cl{\bf THE ``EXTRA-SOLAR GIANT PLANETS" ARE BROWN DWARFS}
\vs\vs\vs
\cl{Robert L. Kurucz}
\cl{Harvard-Smithsonian Center for Astrophysics}
\vs
\vs
\cl{May 30, 2001}
\vs
\vs
\vs\vs\vs
\vfill
\eject
\cl{\bf THE ``EXTRA-SOLAR GIANT PLANETS" ARE BROWN DWARFS}
\vs
\cl{Robert L. Kurucz}
\vs
\cl{Harvard-Smithsonian Center for Astrophysics}
\cl{60 Garden St, Cambridge, MA 02138}
\vs
\vs
After an M, K, G, or F star forms, it magnetically compresses the infall  

dregs to produce a close-in brown dwarf.

\vs
\vs
 
     A new M, K, G, or F star is rapidly rotating and magnetic.  It rotates at, say, 
60 km/s, or once per day.  Figure 1 shows the present solar magnetic field when 
rotation has decayed to 2 km/s or once a month.  Figure 2 shows the same view
of the solar field 4.5 billion years ago.  The lightly-wound field now 
originally had many windings between the star and the ``Mercury" wall. Figures 
3 and 4 show the side views.  Weak jets of rejected material blow from the poles.  
The remaining structure is a magnetic compressor.  Neutral material is drawn 
toward the star by gravity from above and below the disk.   These 
are only the dregs of the cloud as the star formation process shuts down.    
Flares, reconnections, etc. ionize the incoming neutral material so that it is 
slowed and compressed and caught by the field lines.  It is also compressed 
from behind by additional infall.  Flares hammer and distort the  field lines 
to orders of magnitude greater effect than now, Figure 5.  These distortions 
prevent material from falling inward.  The infall forms a highly compressed, 
self-gravitating double torus above and below the equator.  The tori are very 
unstable.  A large flare triggers their rapid collapse into one or two brown 
dwarfs orbiting inside the orbit of ``Mercury".  One brown dwarf can have a 
stable orbit.  If there are two brown dwarfs, one or both are ejected to wider 
orbits or are accreted by the star.  If the dregs are insufficient to form
a brown dwarf, they are eventually eroded away by the wind.
   
The brown dwarf or dwarfs have low abundances because the magnetic field 
separates elements with high first ionization potential that are neutral 
from elements with low first ionization potential that are ionized.
Here is the list of elements with their ionization potentials in cm$^{-1}$
that are not ionized by Lyman $\alpha$ or $\beta$:  H 109678, He 198310, N 117225, 
O 109837, F 140524, Ne 173929, Cl 104591, Ar 127109, Kr 112914.
As neutrals, these elements pass deep into the magnetic trap until they are 
suddenly ionized by flares.  They are unable to get out and they join the 
trapped plasma that eventually forms a brown dwarf.

All other elements have low ionization potentials so are ionized by the 
photospheric and chromospheric radiation from the protostar.  If they fall
toward the protostar, they spiral on the divergent magnetic field lines and 
are expelled far from the star itself.  The easily ionized elements have 
low abundances in the final brown dwarf.
\vfill
\eject
As the surface of the brown dwarf cools, molecules form such as H$_{2}$, NH, OH,
HF, HCl, N$_{2}$, O$_{2}$, NO, H$_{2}$O, O$_{3}$, N$_{2}$O, NO$_{2}$, H$_{2}^{+}$, 
H$_{3}^{+}$, dimers, etc.

If the brown dwarf is accreted by the star, the star will have reduced apparent 
abundances depending on the relative masses of the brown dwarf and 
of the convective zone of the star.

Of course, the total abundances integrated through the whole star in nearby 
Population I stars less than, say, 4 billion years old, are equal to, or 
greater than, solar.

The bodies that are now being discovered called ``extra-solar giant planets" are
not planets but brown dwarfs.  Brown dwarfs formed by other mechanisms could have 
high abundances.

\vs
I hope that someone capable of computing this scenario will investigate
it in detail.     

\vfill
\end